\documentclass[aps,prl,floatfix,twocolumn,,superscriptaddress,showpacs]{revtex4}
\usepackage{amsfonts}
\usepackage{amsmath}
\usepackage{amssymb}
\usepackage{graphicx}
\usepackage{dcolumn}% Align table columns on decimal point
\usepackage{bm}% bold math

\begin{document}

\title{ Thermodynamic modelling of phase separation in manganites}
\author{J. Sacanell}
\affiliation{Departamento de F\'{\i}sica, Comisi\'{o}n Nacional de Energ\'{\i}a At\'{o}%
mica, Av. Gral Paz 1499 (1650) San Mart\'{\i}n, Buenos Aires, Argentina}
\author{F. Parisi}
\affiliation{Departamento de F\'{\i}sica, Comisi\'{o}n Nacional de Energ\'{\i}a At\'{o}%
mica, Av. Gral Paz 1499 (1650) San Mart\'{\i}n, Buenos Aires, Argentina}
\affiliation{Escuela de Ciencia y Tecnolog\'{\i}a, UNSAM, Alem 3901, San Mart\'{\i}n,
Buenos Aires, Argentina}
\author{J.C.P. Campoy}
\affiliation{Instituto de F\'{\i}sica Gleb Wataghin, UNICAMP, Campinas, SP
13083-970,Brazil}
\author{L. Ghivelder}
\affiliation{Instituto de F\'{\i}sica, Universidade Federal do Rio de Janeiro, C.P.
68528, Rio de Janeiro, RJ 21941-972, Brazil}
\pacs{75.30.Kz, 72.80.Ga, 71.23.Cq}

\begin{abstract}
We present a phenomenological model based on the thermodynamics of
the phase separated state of manganites, accounting for its static
and dynamic properties. Through calorimetric measurements on
La$_{0.225}$Pr$_{0.40}$Ca$ _{0.375}$MnO$_{3}$ the low temperature
free energies of the coexisting ferromagnetic and charge ordered
phases are evaluated. The phase separated state is modeled by free
energy densities uniformly spread over the sample volume. The
calculations contemplate the out of equilibrium features of the
coexisting phase regime, to allow a comparison between magnetic
measurements and the predictions of the model. A phase diagram
including the static and dynamic properties of the system is
constructed, showing the existence of blocked and unblocked
regimes which are characteristics of the phase separated state in
manganites.
\end{abstract}

\date{\today}
\maketitle

\section{I. Introduction}

The issue of phase separation is currently one of the main topics of
research in strongly correlated electron systems.\cite{Nanoscale} Phase
separation (PS) fully develops in various manganese binary oxides, but there
are also evidences of the key role played by clustered states in high Tc
superconductors.\cite{DagottoJMMM} Yet, after several years of intense
experimental and theoretical research in this area, the true nature of the
phase separated state observed in manganites is still controversial. Some
phenomenological models points to the strain between the coexisting phases
as the main reason for the appearance of phase separated states.\cite%
{Self,Nature04} In addition, there is a lot of theoretical evidence in favor
of the role of intrinsic disorder in the stabilization of the phase
separated state. These models are based mainly in the double exchange
theory, with a fundamental role played by electron-phonon coupling.\cite%
{Khomskiieulet,Motome,DagottoPRB04} Khomskii and co workers have also
pointed out the tendency to PS of double exchange Hamiltonians when elastic
interactions are included.\cite{Kagan99}

The presence of quenched disorder can lead to a rough landscape of the free
energy densities, triggering the formation of clustered states, which are
induced by phase competition.\cite{DagottoJMMM} Moreo et al. obtained phase
separated states in a Monte-Carlo simulation of a random-field Ising model
when disorder is included in the coupling and exchange interactions.\cite%
{MoreoPRL} Similar results were obtained by Burgy and coworkers, using a
uniform distribution of exchange interactions.\cite{Burgy01} As shown for
the case of first order transitions, quenched impurities can lead, under
certain circumstances, to a spread of local transition temperatures (where
local means over length scales of the order of the correlation length)
leading to the appearance of a clustered state with the consequent rounding
of the first order transition.\cite{Imry} Typical mechanisms to include
disorder are chemical\cite{NatureUeh} and structural.\cite%
{Podzorov,Controled}

Despite the intense effort towards a microscopic understanding of the
manganites,\cite{Carvaj,ThomasPRL} many macroscopic features of the phase
separated state, including its thermodynamic properties and dynamic
behavior, still remains to be studied in greater detail. One of the most
interesting features of the phase separated state is the entwining between
its dynamic and static properties. Some of the phase separated manganites
display slow relaxation features that hide experimentally the real
equilibrium thermodynamic state of the system.\cite{Novel} This is why the
construction of phase diagrams is currently focused on the dynamic
properties of the phase separated state, with regions of the phase diagrams
nominated as \textquotedblleft frozen\textquotedblright\ or
\textquotedblleft dynamic\textquotedblright\ PS,\cite{Dynamic}
\textquotedblleft strain glass\textquotedblright\ or \textquotedblleft
strain liquid\textquotedblright \cite{reentrant}\ or, directly, ascribed as
spin glass phases.\cite{PRLglass,PRLMathieu}

Among the challenging issues that have not yet been addressed is an
understanding of the phase separated state in terms of its thermodynamic
properties. An analysis of the behavior of phase separated systems based on
the probable free energies functional has been schematically realized,\cite%
{Ueharaeulet,Babush} but without the corresponding measurements supporting
the proposed scenario. In the present study we perform an attempt to
construct the thermodynamic potentials of the FM and non-FM phases of a PS
manganite, through calorimetric and magnetic measurements. The experiments
were carried out in a polycrystalline sample of La$_{5/8-y}$Pr$_{y}$Ca$%
_{3/8} $MnO$_{3}$ ($y=0.4$), a prototypical phase separated system
in which the effects of the substitution of La by Pr produces an
overwhelming effect on its physical properties.\cite{NatureUeh} We
took advantage of the fact that in the mentioned compound
homogeneous phases can be obtained at low temperatures in long
time metastable states, which allow us to measure separately the
specific heat of each phase, CO or FM, in the low temperature
region (between 2K and 60K). With this data it is possible to
write an expression for the difference between the Gibbs energies
of the homogeneous phases as a function of temperature and
magnetic field. The needed constants to link the thermodynamic
potentials of both phases were obtained from indirect measurements
based on the static and dynamic behavior of the system. The phase
separated state is modeled through the hypothesis that the free
energy densities are spread over the sample volume, and that its
non-equilibrium features are governed by a hierarchical
cooperative dynamic. Within this framework it is possible to
construct a phenomenological expression for the free energy of the
phase separated state based on experimental data, which is able to
describe consistently the behavior of the system as a function of
temperature and applied magnetic field.

\section{II. Experimental Details}

The measurements were made on a polycrystalline sample of La$_{0.225}$Pr$%
_{0.40}$Ca$_{0.375}$MnO$_{3}$. Details of material preparation were
previously published.\cite{Dynamic} Both magnetization and specific heat
results were obtained with a Quantum Design PPMS system. Magnetization data
was measured with an extraction magnetometer, as a function of temperature,
applied magnetic field, and elapsed time. All temperature dependent data was
measured with a cooling and warming rate of 0.8 K/min. Specific heat data
was measured with a relaxation method, between 2 and 60 K.

\section{III. Results and Discussion}
\begin{center}
\begin{figure}
  % Requires \usepackage{graphicx}
  \includegraphics[height=\columnwidth,angle=-90]{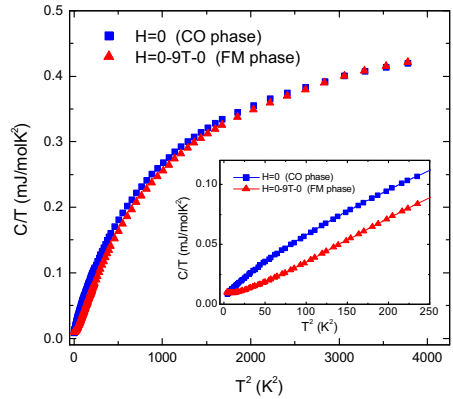}
  \caption{(color online). Specific heat of La$_{0.225}$Pr$_{0.40}$Ca$_{0.375}$%
MnO$_{3}$ as a function of temperature for $H$=0 in the CO phase
(squares) after ZFC, and in the FM phase (triangles) after ZFC to
2K following of a 0-9T-0 field sweep. The inset shows an
enlargement of the low temperature data.}\label{fig1}
\end{figure}
\end{center}

Following the $H-T$ phase diagram of the compound,\cite{Dynamic,reentrant}
after zero field cooling (ZFC) the sample reaches the low temperature state
mainly in the CO phase, a state that has been described as \textquotedblleft
frozen PS\textquotedblright \cite{Dynamic} or \textquotedblleft strained
glass\textquotedblright \cite{reentrant} This frozen state can be released
by the application of a moderate magnetic field ($H$=2.2T),\cite{Abrupt}
above which the compound transforms into the FM phase in an abrupt
metamagnetic transition.\cite{Schiffer,Fisher,Rana} After this step
transition the sample remains in this homogeneous FM state until a
temperature around 70 K, even after the magnetic field is removed. These
facts were used to perform the measurements of the specific heat $c^{\alpha
} $ of each phase ($\alpha =$CO or FM) between 2K and 60K, each one
considered as homogeneous in this temperature range under specific field
conditions. In Fig. 1 the plot of $c/T$ vs $T^{2}$ is shown for measurements
performed while warming after ZFC with different procedures: under zero
field (CO phase), and under different fields $H$ after a field sweep 0-9T-$H$%
, for $H$=0, 1 and 2 T (FM phase). The data of the CO phase and that of the
FM are clearly distinguished. Also, the results obtained for the FM phase
for the fields employed are practically identical, which is a signature that
the FM phase obtained after the application of 9T remains homogeneous until
the highest temperature investigated. Besides, the fact that the specific
heats of \ the FM phase are almost independent of $H$ is a signal that, in
the range of temperature investigated, there is no significant field
dependent contribution to the entropy of the FM phase, indicating that the
magnetization is saturated for all fields. The data obtained was adjusted
using standard models for CO and FM phases.\cite{Hardy} The small upturn
observed at low temperatures corresponds to the onset of the ordering of the
magnetic momentum of the Pr atoms.\cite{Smolya}

The thermodynamic Gibbs potential $g$ of each phase may be written as $%
g_{^{0}}^{\alpha }(T,H)=E^{\alpha }(T,H)-TS^{\alpha }(T,H)$ where the
superscript $\alpha $ indicates the phase (CO or FM), $E$ and $S$ are the
enthalpy and the entropy respectively. From the specific heat data we could
construct both E and S in the usual way:\cite{Landau}
\begin{equation}
E^{\alpha }(T,H)=E_{0}^{\alpha }+\int\limits_{T_{0}}^{T}c^{\alpha }(T)dT-MH
\end{equation}

\begin{equation}
S^{\alpha }(T,H)=S_{0}^{\alpha }+\int\limits_{T_{0}}^{T}\frac{c^{\alpha }(T)%
}{T}dT
\end{equation}%
where $M$ is the magnetization of the phase (we assume $M$=0 for the CO
phase) and $T_{0}$=2K is the lowest temperature reached in the measurements.
As the specific heat of the FM phase was found almost independent of $H$, we
consider the Zeeman term in Eq. (1) as the only dependence of the Gibbs
potentials of the FM phase with $H$, so that $%
g_{^{0}}^{FM}(T,H)=g_{^{0}}^{FM}(T,0)-MH$. No dependence with $H$ is
considered for the CO phase, since there is no way to perform the
measurements on the CO phase under an applied $H$ due to its instability
against the application of $H$ in the temperature upturn. The terms $%
E_{0}^{\alpha }$ and $S_{0}^{\alpha }$\ are respectively the values of the
enthalpy and the entropy at the initial temperature $T_{0}$. Since we are
interested in the energy difference between the phases involved, we have
taken $E_{0}^{FM}=0$ and $S_{0}^{FM}=0$ as reference values, remaining $%
E_{0}^{CO}$ and $S_{0}^{CO}$as the constant to be determined.

In order to determine $E_{0}^{CO}$ we followed the previously published
experimental data in relation to the abrupt magnetic transition from the CO
to the FM phase, which happens at low temperatures (below 6K), under a
magnetic field of around 2.2 T [Ref. \cite{Abrupt}]. This transition is
accompanied by a sudden increase of the sample temperature, which reaches a
value around 30K after the transition. Due to the velocity of the process,
it is plausible to consider that the enthalpy remains constant at the
transition point, so that the following relation is fulfilled:

\begin{equation}
E_{0}^{CO}=\int\limits_{2K}^{30K}c^{FM}(T)dT-M_{sat}H
\end{equation}%
where $M_{sat}$ is the magnetization of saturation of the FM phase ($M_{sat}$%
=3.67$\mu _{B}$ /Mn=20.5 J/molT) and $H$=2.2T. This calculation yields $%
E_{0}^{CO}=28.3J/mol$, indicating that the homogeneous FM phase has lower
free energy at low temperatures even for $H$=0, as suggested in Ref. \cite%
{Dynamic}.

In order to fully construct the thermodynamic potentials we have
to determine the remaining constant $S_{0}^{CO}$ which, at this
point, is what controls the transition temperature between the
homogeneous phases, providing it is a positive quantity as
expected for the difference of entropy between the FM and CO
phases due to the excess of configurational entropy of the
latter.\cite{Ueharaeulet} In Fig. 2 the obtained thermodynamic
potentials of the homogeneous phases are displayed, assuming a
value $S_{0}^{CO}=0.65J/molK$, this value was obtained by adjusting the $%
M(H) $ curve of Fig. 4b, as explained below. The plot indicates that the
homogeneous FM state has a lower energy than the CO state for all fields at
low temperatures, while the CO phase is the stable state at high
temperatures, with field-dependent transition temperatures ranging from 30K
for $H$=0 to 60K for $H$=2T.

\begin{center}
\begin{figure}
  % Requires \usepackage{graphicx}
  \includegraphics[height=\columnwidth,angle=-90]{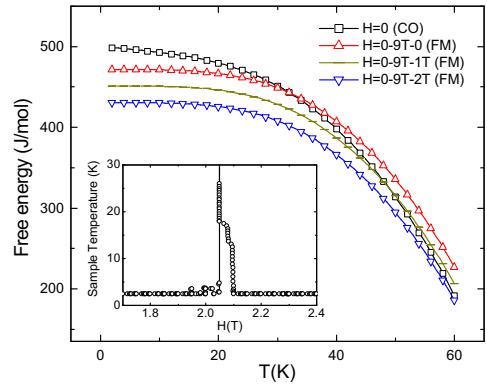}
  \caption{(color online). Free energies as a function of $T$ of the homogeneous
phases calculated through Eqs. (1) and (2) for different field
sweeps. Inset: temperature of the sample as a function of $H$ when
the step transition occurs (taken from Ref. \cite{Dynamic}); the
base temperature is 2.5 K.}\label{fig2}
\end{figure}
\end{center}

The above presented results give us an insight on the behavior of
the system under the hypothesis that no phase separation occurs,
i.e., describes the thermodynamics of the homogeneous equilibrium
phases. In addition, in order to obtain a phase separated state
from the thermodynamic data an appropriate modelling needs to
include a priori the existence of the phase separated state.
However, one needs to be careful when comparing the predictions of
the model with experimental results. It is well established \cite%
{Dynamic,reentrant} that the phase separated state is characterized by a
slow dynamics, which implies that equilibrium is hardly reached in
laboratory times. The equilibrium properties must be linked with the
measured data and therefore a dynamic treatment is needed.

In the discussion that follows we perform a qualitative analysis within a
framework where both static and dynamic properties are treated on a
phenomenological basis. It is well known that the physical properties of the
LPCMO system changes dramatically near $y$=0.32, \cite{NatureUeh} revealing
the extreme sensitivity of the system to small variations in the mean atomic
radius of the perovskite A-site. This can be due to the effects that
quenched disorder introduced by chemical substitution has on the local
properties, \cite{MoreoPRL} or else to the role played by "martensitic-type"
accommodation strains originated by the volume differences between the FM
and CO unit cells. \cite{Ueharaeulet,Podzorov} On one hand, the inclusion of
disorder in a random-field Ising model leads to a spatially inhomogeneous
transition temperature, from the paramagnetic disordered phase to the
"ordered" phase, characterized by the appearance of clustered states.\cite%
{Burgy01,Aliaga} This fact implies a spatial dependence of the quadratic
coefficient in a Landau-type expansion. On the other hand, strain induced by
the shape-constrained transformation between the CO and the FM phases could
lead to the phase separated state through the frustration of long-range
interactions. \cite{Ueharaeulet,Podzorov} In the latter view, the properties
of a specific compound are governed by an "effective" Pr concentration,
which controls the capability of the system to accommodate the anisotropic
strain.\cite{Ueharaeulet} These two alternative pictures are not mutually
exclusive; the clustered states induced by disorder are enhanced if
correlated disorder is included in the model,\cite{Burgy04} a feature that
mimics the cooperative effects of the Jahn-Teller distortion, in a similar
way as elastic interactions are able to induce long scale phase separation
in phenomenological models \cite{Nature04} (in this last case a renormalized
fourth order term is the responsible for the introduction of spatial
inhomogeneities). Additionally, local variations of the atomic composition
can couple with the anisotropic strain triggering the formation of the phase
separated state.\cite{Ueharaeulet}

These facts can be qualitatively described introducing non-uniform free
energy densities for the CO and FM phases. The simplest form is a uniform
distribution of these densities over the sample volume, with mean values
equal to the free energy of the homogeneous phases. Within the hypothesis
that precursor effects of phase separation are due to variations of the
local composition, we follow Imry and Wortis theory \cite{Imry} describing
the effects of disorder on a system displaying a first order transition in
order to estimate the width of the free energies distributions. Following
their ideas, and considering that disorder affects mainly the free energy
density of the FM phase, ($T_{co}$ is nearly constant as a function of Pr
content $y$ [Ref. \cite{NatureUeh}]) we can write an expression for the
local free energy density $g^{FM}(\Delta y)$ depending on the fluctuations
of composition $\Delta y$.

\begin{equation}
g^{FM}(\Delta y)=g_{0}^{FM}+(S^{CO}-S^{FM})\frac{dT_{C}}{dy}\Delta y
\end{equation}%
where $\Delta y$ is taken over length scales comparables with the
correlation length, which is around 1 nm for microscopic clusters in Pr$%
_{07} $Ca$_{03}$MnO$_{3}$. \cite{Radaelli} With the Gaussian
distribution proposed in Ref \cite{Imry} $\Delta y$ can be as high
as 0.1 over nanometer length scales; the development of micrometer
sized domains would need of the consideration of elastic
interactions. For the sake of simplicity we take a uniform
distribution for $\Delta y$ over the sample volume. With these
assumptions, the free energy density of the FM phase as a function
of $x$ (the volume coordinate is normalized to 1) is written as:

\begin{equation}
g^{FM}(x)=g_{0}^{FM}-\delta (1/2-x)
\end{equation}%
where $\delta $\ is the parameter controlling the width of the free energy
functional, which can be estimated from Eq. (4) as $\delta \approx
(S_{co}-S_{FM})\frac{dT_{C}}{dy}y/2\approx (70K)\,S_{^{0}}^{co}$, taking
into account that $\frac{dT_{C}}{dy}\approx -350K$ [Ref.\cite{NatureUeh}]
and assuming $\Delta y\approx \frac{y}{2}(1/2-x).$ In this way, the
equilibrium FM fraction $x_{eq}$ at the given $T$ and $H$ is obtained as:

\begin{equation}
x_{eq}=\frac{g_{0}^{CO}(T)-g_{0}^{FM}(T,H)+\frac{1}{2}\delta }{\delta }
\end{equation}%
This expression could be used for determining the parameters $S_{co}$ and $%
\delta $, using for instance the $M(H)$ data at different temperatures.
However, as stated before, the global behavior of the system at low
temperatures is characterized by out-of-equilibrium features, so the true
values for $x_{eq}$ are not easily accessible experimentally.

In order to circumvent the fact that the thermodynamic equilibrium state is
not reached experimentally, an alternative approach is to consider that the
response of the system within the phase separated regime as a function of
temperature can be qualitatively described within a model of cooperative
hierarchical dynamics, using an activated functional form with
state-dependent energy barriers.\cite{Dynamic} The time evolution of the FM
fraction $x$ is given by:

\begin{equation}
\frac{dx}{dt}=\frac{(x_{eq}-x)}{|x_{eq}-x|}v_{0}e^{-\frac{U}{T|x_{eq}-x|}}
\end{equation}%
where $\nu _{0}$ represents a fixed relaxation rate and $U$ is a (field
dependent) energy barrier scale. This model is similar to that employed to
describe vortex dynamics in high $T_{C}$ superconductors\cite{Blatter} and
is based in dynamic scaling for systems with logarithmic relaxations.\cite%
{Labarta} The inbreeding between the dynamic behavior and the equilibrium
state of the system is given by the functional form of the effective energy
barriers $\frac{U(H)}{|x_{eq}-x|}$, which diverge as the system approaches
equilibrium. This fact represents the main difference with respect to a pure
superparamagnetic behavior, and predicts the existence of state-depending
blocking temperatures $T_{B}(H,x_{B})$ at which, for any given FM fraction $%
x_{B}$ , the system becomes blocked, in the sense that the velocity of
change of the FM fraction is lower than, for instance, the detectable
velocity $\nu _{\exp }$, estimated as $\approx $ 10$^{-6}s^{-1}$ for a
conventional data measurement that takes 30 sec.\cite{Dynamic}. This gives
the following relation for $T_{B}(H,x_{B})$:

\begin{equation}
T_{B}(H,x_{B})\leq \frac{U(H)}{|x_{eq}-x_{B}|\ln (v_{0}/v_{\exp })}
\end{equation}

Through Eq. (8) it is possible to obtain an experimental estimation of the
factor $U(H)/\ln (v_{0}/v_{\exp })$ which governs the interplay between the
dynamics of the system and the measurement procedure. Figure 3 shows $M(H)$
data for selected temperatures, and the experimentally obtained values of $%
T_{B}(H,x_{B}=\frac{1}{2})$\ as a function of $H$, determined through low
temperatures measurements between 6 and 32 K. The values of $H$ are those at
which the system reaches the state with $x_{B}=0.5$. The main assumption is
that each point of the $M(H)$ curve corresponds to a blocked state
compatible with the measurement procedure. As indicated in the inset of
Figure 3 the relation $\frac{1}{2}T_{B}(H,\frac{1}{2})\approx 25H^{-\beta }$
holds, with $\beta \approx 2.33$. We will show later that the equilibrium
state at low temperatures is fully FM, so Eq. (8) is a direct measurement of
the field dependence of the factor $U(H)/\ln (v_{0}/v_{\exp })=\frac{1}{2}%
T_{B}(H,\frac{1}{2})$. Within the dynamic model adopted, this
factor is independent of the particular value of $x_{B}$ choosen
for its determination. With the assumption that this relation
holds in the whole temperature range where phase separation
occurs, we can write a simple equation relating the equilibrium
state with the experimentally accessible state

\begin{equation}
\left\vert x_{eq}(T,H)-x_{B}(T,H)\right\vert \leq \frac{U(H)/\ln
(v_{0}/v_{\exp })}{T}
\end{equation}%
Through this equation it is possible to make the link between the
equilibrium state of the system, which can be obtained through the free
energies functional, and the experimental data obtained through both $M(H)$
and $M(T)$ measurements.

\begin{center}
\begin{figure}
  % Requires \usepackage{graphicx}
  \includegraphics[height=\columnwidth,angle=-90]{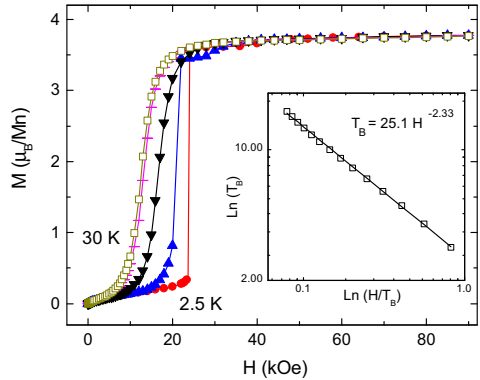}
  \caption{(color online). M(H) data at selected temperatures: 2.5, 8, 15,
25, and 30 K. Inset: log-log plot of $T_{B}$ versus $H/T_{B}$,
where $H$ is that for which $M$ is half of its saturation value.
}\label{fig3}
\end{figure}
\end{center}

The upper panel of Fig. 4 sketches the temperature evolution of
$x_{eq}$ for different magnetic fields, obtained from Eq. (6) with
$S_{0}^{CO}=0.65J/molK$ and $\delta \approx
(70K)\,S_{^{0}}^{co}=45.5J/mol$. It shows that, with the set of
parameters employed, the low temperature state of the system is
fully FM for moderate fields. However, the accessible FM fraction
after a ZFC procedure is small for $H$ $<$ 2 T, due to the weight
of the blocking term in Eq. (9). This is why, besides the fact
that the difference between the free energies of the CO and the FM
states increases as temperature is lowered, the magnetic field
needed to induce the CO-FM transition also increases, a fact that
at first sight could be interpreted as a reentrance of the CO
state. As the temperature is raised above 25K the influence of the
blocking term decreases; in this temperature region the main
factor determining the field needed to induce de CO-FM transition
lies in the field dependence of the equilibrium fraction.

\begin{center}
\begin{figure}
  % Requires \usepackage{graphicx}
  \includegraphics[width=\columnwidth]{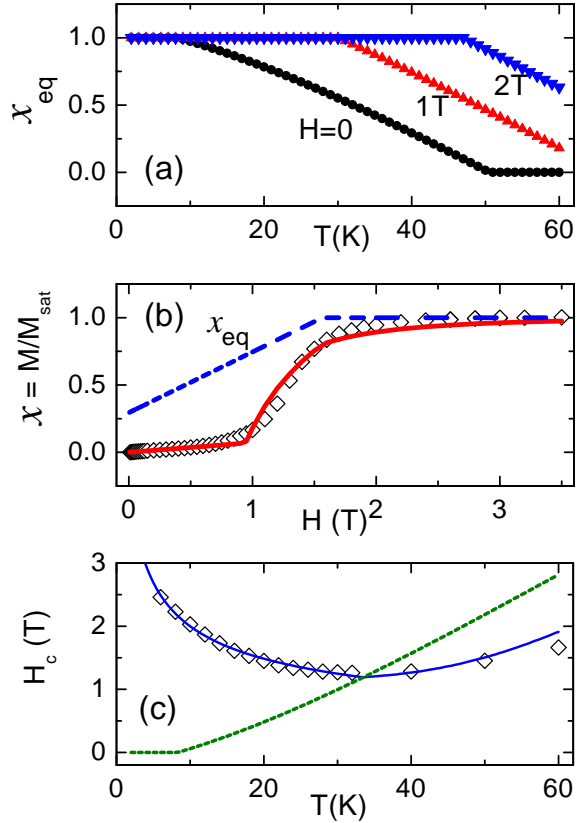}
  \caption{(color online). a) Temperature dependence of the equilibrium FM
fraction, $x_{eq}$, for the indicated fields. b) FM fraction $x$
as a function of $H$ obtained through $M(H)$ data (open symbols),
and calculated with Eq. (9) (solid line) for $T=40$K. The FM
equilibrium fraction $x_{eq}$ obtained from Eq. (6) is also
displayed (dashed line). c) Field needed to make the system half
FM, as a function of $T$, from $M(H)$ measurements (open symbols)
and calculated through Eq. (9) (solid line). The temperature
dependence of the field needed to make the equilibrium state of
the system fully FM is also displayed (dashed line).}\label{fig4}
\end{figure}
\end{center}

The middle and lower panels of Figure 4 show the comparison of experimental
data with the results obtained through Eq. (9). In Fig. 4b, $M(H)$ data at
40 K normalized by its saturation value and the corresponding calculated
values are displayed, showing the good agreement between them. Figure 4c
shows measured and calculated values of the field $H_{c}$ at which $%
x_{B}=0.5 $, as a function of temperature; this is the field needed to make
half the system FM. Also shown is the temperature dependence of the field
needed to make the equilibrium state of the system fully FM. As can be seen,
the calculated curve for $H_{c}$ reproduces the experimental behavior, with
a minimum around 30K. This minimum signals the crossover from the blocked
regime at low temperatures (frozen PS) to the coexistence regime at higher
temperatures (dynamic PS). In the frozen PS regime the stable state of the
system is homogeneous FM for all fields needed to induce the growth of the
FM phase to a value $x=1/2$; the presence of the CO phase is only explained
by the slow growing of the FM phase against the unstable CO due to the
energy barriers. In the dynamic PS regime, the influence of the blocking
term diminished, the equilibrium state is true PS for moderate fields, and
the effect of $H$ is mainly to unbalance the amount of the coexisting phases.

Figure 5 shows the $x-T$ state diagram obtained from Eqs. (6) and (9) for a
field $H=1.3$T, which displays both dynamic and static properties of the
model system. A line in the phase diagram divides it in two major regions,
depending on the equilibrium FM fraction $x_{eq}$. A point above the line
indicates that the system has an excess of FM phase; below the line the
state of the system is characterized by the presence of metastable CO
regions. Each of these regions is in turn divided in two other, labeled $%
dynamic$ $CO$ or $dynamic$ $FM$, indicating that if the system is
in a state within this region is able to evolve toward equilibrium
within the measurement time. The regions labeled as $frozen$
indicate that the system is blocked, and no evolution is expected
within the time window of the experiment. Data points
$x(T,1.3$T$)$ obtained from $M(H)$ and $M(T)$ measurements are
also showed. The $M(H)$ data is obtained in a field sweep after
ZFC to the target temperature, and the $M(T)$ in a field warming
run after ZFC to 2K. The data extracted from the $M(H)$
measurements gives information on the $x$ values for which the
system becomes blocked at each temperature, for the specific field
and for the characteristic measuring time, indicating the frontier
between the dynamic and frozen FM regions. The data obtained from
$M(T)$ measurements coincides with that of $M(H)$ in the low
temperature region, where $\partial x_{B}(H,T)/\partial T>0$. As
the temperature is increased, the system gets into a region for
which $\partial x_{B}(H,T)/\partial T<0$ with a FM fraction
greater than the lower limit for $x_{B}$, so it remains blocked
without changes in the magnetic response, a fact characterized by
the plateau observed in $M(T)$ for the high temperature region.
This non-equilibrium phase diagram was constructed for the
particular measurement procedure employed for the acquisition of
$M(H)$ and $M(T)$ data. A modification in this procedure (for
instance, by changing the time spend at each measured point) will
result in a change in the factor $\nu _{\exp }$, with the
consequent change of the boundaries in the phase diagram. For
example, the effect of increasing the measured time by one order
of magnitude is shown by dashed lines in the phase diagram of Fig.
5.

\begin{center}
\begin{figure}
  % Requires \usepackage{graphicx}
  \includegraphics[width=\columnwidth]{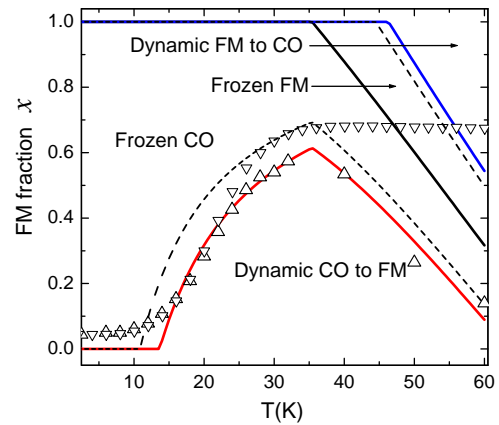}
  \caption{(color online). State diagram of the system in the $x-T$ plane, for $%
H=1.3$T, resulting from Eq. (9). Regions labeled dynamic indicate
that the system can evolve in the measured time. Those labeled
frozen indicates that the system is blocked. Data obtained from
$M(H)$ measurements at different temperatures ( up triangles) and
from $M(T)$ at $H$=1.3 T after ZFC (down triangles) are also
shown. The dashed lines show the new phase diagram boundaries if
the measuring time is increased by one order of
magnitude.}\label{fig5}
\end{figure}
\end{center}

\section{IV. Conclusions}

In conclusion, we presented a thermodynamic phenomenological model for a
global description of the phase separated state of manganites. The
construction starts with the calculation of the free energies of the
homogeneous FM and CO phases. The free energies obtained turned out to be
very close in value: the difference was of the order of the magnetic energy
for intermediate fields in the whole temperature range investigated. The
phase separated state is introduced by considering a uniform spread of the
free energy density of the FM phase, and the dynamic behavior is included
within a scenario in which the evolution of the system is determined through
a cooperative hierarchical dynamic with diverging energy barriers as the
system approaches equilibrium. The main success of the model is to provide
an understanding of the response of the phase separated state when both
temperature and magnetic field are varied, being able to reproduce the
dynamic and static properties of the system under study. The same
methodology can be also applied to other compounds sharing similar phase
diagrams \cite{Radaelli,Hervieu} and properties, especially those displaying
abrupt field induced transitions at low temperatures.\cite%
{Schiffer,Fisher,Rana} The key factors to determine the free energies of the
homogeneous phases is the possibility to measure the specific heat of each
phase separately, taking advantage of the existence of blocked states, and
the measurement of the temperature reached by the compound under study after
the CO-FM abrupt transition at low temperature. This last value and the
field at which the abrupt transition occurs are the key parameters to
determine the homogeneous ground state of the system at zero applied field.

This work was partially supported by Fundaci\'{o}n Antorchas (Argentina) and
CAPES, CNPq, and FAPERJ (Brazil).

\end{document}